\documentclass[sigconf,nonacm]{aamas} 
\usepackage{balance} % for balancing columns on the final page
\usepackage{cleveref} % for better references
\usepackage{xspace}
\usepackage[inline]{enumitem}

\newtheorem{opportunity}{Opportunity}

\setcopyright{ifaamas}
\acmConference[AAMAS '24]{Proc.\@ of the 23rd International Conference
on Autonomous Agents and Multiagent Systems (AAMAS 2024)}{May 6 -- 10, 2024}
{Auckland, New Zealand}{N.~Alechina, V.~Dignum, M.~Dastani, J.S.~Sichman (eds.)}
\copyrightyear{2024}
\acmYear{2024}
\acmDOI{}
\acmPrice{}
\acmISBN{}

\acmSubmissionID{}

\title{Elections in the Post-Quantum Era:\\ Is the Complexity Shield Strong Enough?}
\subtitle{A Call for New Axiomatic, Hardness, and Operation-Research Results for Voting Theory}

\author{Šimon Schierreich}
\affiliation{
	\institution{Czech Technical University in Prague}
	\city{Prague}
	\country{Czechia}%
}
\email{schiesim@fit.cvut.cz}

\begin{abstract}
	The election, a cornerstone of democracy, is one of the best-recogni-zable symbols of democratic governance. Voters' confidence in elections is essential, and these days, we can watch practically in live broadcast what consequences distrust in the fairness of elections may have. From the times of the celebrated Gibbard–Satterthwaite theorem, it is well-known in the social-choice community that most voting systems are vulnerable to the efforts of various players to influence elections. Luckily for us, computing such influence to affect election outcomes is a hard problem from the computational complexity perspective. This intractability is regarded as a ``complexity shield'' that secures voting rules against this malicious behavior.
	
    In this work, we consider quantum computers to be a new threat to the complexity shield described above, as they break out of standard computing paradigms and unlock additional computational resources. To this end, we provide an overview of possible attacks on election, discuss the abilities of quantum computing, and chart possible directions for future research in this area.
\end{abstract}


\newcommand{\NP}{\textsf{NP}\xspace}
\newcommand{\NPh}{\NP-hard\xspace}

\usepackage{cleveref}
\begin{document}
	
%%% The following commands remove the headers in your paper. For final 
%%% papers, these will be inserted during the pagination process.

\pagestyle{fancy}
\fancyhead{}

%%% The next command prints the information defined in the preamble.

\maketitle 

%%%%%%%%%%%%%%%%%%%%%%%%%%%%%%%%%%%%%%%%%%%%%%%%%%%%%%%%%%%%%%%%%%%%%%%%

\section{Introduction}

Elections and voting are widely regarded as crucial and perhaps the most visible symbols of democratic governance, although alternative mechanisms for selecting representative bodies exist (see, e.g., the literature on liquid democracy~\cite{BlumZ2016,ChristoffG2017,GolzKMP2021,KahngMP2021} or recent works of Flanigan et al.~\cite{FlaniganGGHP2021,FlaniganLPW2024}). Ensuring that citizens trust the election results is fundamental for the survival of democracy itself.
To illustrate the potential causes of distrust in election results, consider an episode from the not-so-distant past in a country that is often referred to as ``the largest democracy in the world''.

\begin{quote}
On 6 January 2021, at 2:13 p.m., the United States Capitol was officially breached. The Secret Service had to evacuate senators and representatives. A mob of protesters broke into the building to thwart the counting of electoral college votes to formalize the victory of the new president.
\end{quote}

Why did they do so? They were convinced that the presidential election was a fraud and that President Biden was elected only due to manipulations. According to a survey by Axios-Momentive~\cite{Richardson2022}, in January 2022, there were still more than 40\% Americans convinced that the new president did not win legitimately. The distrust of protesters about the outcome of the election led to at least 5 killed and 146 injured during the events~\cite{Heine2021}.

Although it can be discussed what were the main causes of the events described above, categorically dismissing each distrust in election results as baseless would be in contradiction to established scientific knowledge. For decades, it has been widely recognized that every voting rule is susceptible to various forms of malicious behavior. Arguably, the most famous result in this context is the following impossibility theorem and its extensions and generalizations~\cite{MullerS1977,DugganS2000}.
\begin{theorem}[\citet{Gibbard1973,Satterthwaite1975}]\label{thm:GS}
     Every general voting rule defined for at least $3$ candidates and selecting a single winner has at least one of the following properties:
     \begin{enumerate}
         \item is \emph{dictatorship} -- there exists a single fixed voter who always determines the winner, or
         \item is \emph{impose} -- there is at least one candidate who can never win regardless of the voters' opinions, or
         \item is \emph{manipulable} -- voters may declare false ballots and obtain a better result for themselves.
     \end{enumerate}
\end{theorem}

The use of voting theory is not exclusive to real-world elections. With the rise of artificial intelligence and multi-agent systems, we also need formal ways to secure collective decision-making processes in these systems. And voting theory, studied for decades, provides good tools for this purpose~\cite{EphratiR1991,BurkaPSZ2022,RichardsSP2006,FaliszewskiHH2010}.
Why do we need to protect these virtual elections between artificial agents before any form of attack? \citet{FaliszewskiP2010} argue that \emph{software agents have all the patience and computing power necessary to perform complicated analysis} and \emph{are not bound by a moral obligation to act honestly, as in multi-agent systems their goal is to maximize their own utility}.

Fortunately, with the rise of computational social choice, generally dated back to the groundbreaking works of Bartholdi III, Tovey, Trick, and Orlin \cite{BartholdiTT1989a,BartholdiTT1989b,BartholdiO1991,BartholdiTT1992}, a solution how to navigate around the impossibility theorems has appeared. This revelation suggests that the situation is not as desperate as it may seem at first glance. Many studies indicate that the computational complexity of determining how to influence elections for desired outcomes is exceptionally challenging and, in real-life instances, practically impossible.

Therefore, this computational hardness is held as a resilient \emph{complexity shield} that safeguards voting rules from potential malicious interventions. Consequently, we can assume that elections are secure, at least on the level of the voting rule; the security of the ballot collection and vote counting are completely different stories, which are way beyond the scope of this paper.

However, as times change, new opportunities and threats emerge, challenging what was previously thought to be set in stone. Additionally, it is inherent in the scientific method to consistently reevaluate and reexamine established truths. Therefore, in this paper, we discuss the impact of recent advances in quantum computing on our beliefs about the fairness and security of elections and voting rules.

We are aware that our paper discusses a topic at the intersection of multiple different fields, including voting theory, computational complexity, and quantum computing, where at least the last field is still not a very common part of the curriculum of computer science study programs. Nevertheless, we have made diligent efforts to ensure that our work remains accessible to readers without prior knowledge in any of these fields.

\section{How to influence elections?}\label{sec:manipulations}

So far, we have been arguing that all reasonable voting rules are vulnerable to attacks whose purpose is to influence the outcome of the election. However, our definitions of both ``election'' and ``attack'' were very hand-waving. In this section, we give a brief introduction to voting theory and different attacks that are studied in the literature. 

An election typically consists of a set of candidates $\mathcal C$, a set of voters $\mathcal V$, and individual voter preferences represented by \emph{ballot}~$\mathcal B_i$ for each voter $i\in \mathcal V$. These ballots come in various forms, including \emph{approval} ballots (listing approved candidates), \emph{ordinal} ballots (linear orders of candidates), and \emph{cardinal} ballots (assigning numerical values to candidates)~\cite{BramsF2007,LaslierS2010,LacknerS2023,Kim2017}.

Given an election input, the election organizer then employs a (known) voting rule $\mathcal R$ to aggregate voters' ballots and select the winning candidate. While other election models exist, such as perpetual voting~\cite{Lackner2020}, blockchain-based elections~\cite{Grossi2022}, or iterated voting~\cite{LevR2012,LevR2016,GrandiLRVW2013,ObraztsovaMPRJ2015,KavnerX2021}, we focus on this natural and simple model based on ballots and well-defined voting rules.

When an external entity tries to influence an election, its target can be both constructive and destructive. In the former, the attacker's aim may be to ensure their preferred candidate wins, while in the latter, the aim is to thwart a despised candidate's victory. These objectives can be achieved through various strategies. We now introduce three main types of influences studied in the literature~\cite{FaliszewskiR2016,ConitzerW2016}. While our description of each strategy will be with respect to the constructive target, the tweak of their definition for the destructive target should be immediate.

\begin{description}
    \item[Manipulation] involves voters intentionally providing incorrect votes to favor their preferred candidate. As demonstrated by \Cref{thm:GS}, every reasonable voting rule is vulnerable to manipulations.

    \item[Control] occurs when an external entity can influence the participation of candidates or voters. The study of control, particularly from a computational complexity perspective, was pioneered by Bartholdi et al.~\cite{BartholdiTT1992}.

    \item[Bribery,] first studied by Faliszewski et al.~\cite{FaliszewskiHH2009}, involves attackers paying voters for modifications of their ballots rather than deleting candidates or voters. Numerous variants have since emerged, each imposing limits on how ballots can be altered~\cite{ElkindFS2009,Faliszewski2008,KariaMD2023,DeyMN2017}.
\end{description}

In addition to these primary attacks, other manipulative tactics are explored in the literature, including scenarios where abstaining from voting benefits certain groups of voters~\cite{Perez2001,Moulin1988,FelsenthalN2019,BrandtGP2017,BrandtHS2019,BrandtMS2022} and efforts to influence districts where voters submit their ballots~\cite{LewenbergLR2017,LevL2019,EibenFPS2020}. Some works even investigate attacks combining multiple targets~\cite{BoehmerBKL2020}, employing multiple operations~\cite{FaliszewskiHH2011}, or combining generalizations within a single election~\cite{BlazejKS2022}.

\section{The Complexity Shield}\label{sec:shield}

In the late 1980s and early 1990s, Bartholdi III, Tovey, Trick, and Orlin initiated research on the computational complexity of problems within the field of social choice~\cite{BartholdiTT1989a,BartholdiTT1989b,BartholdiO1991,BartholdiTT1992}. Of particular relevance to our current study, Bartholdi III et al.~\cite{BartholdiTT1992,BartholdiTT1989b,BartholdiO1991} introduced the concept of the ``resistance of elections to strategic behavior'' based on the intractability of associated computational problems. They demonstrated that when it is computationally intractable (typically characterized as \textsf{NP}-hard or even harder in terms of the polynomial hierarchy\footnote{For the sake of easiness, we assume that the problems under study are not total-search problems, where the solution is guaranteed to exist but can be computationally hard to find one~\cite{Papadimitriou1994}. 
Nevertheless, even within the computational social choice, such problems naturally appear~\cite{Goldberg2011}.}) to compute a successful influence strategy against a specific voting rule $\mathcal{R}$, that voting rule is considered to be protected by a \emph{complexity shield} against strategic manipulation due to its practical infeasibility.

Which voting rules are vulnerable to the described attacks and which benefits from the complexity shield? Either side contains many representatives of well-known and widely used rules. Because providing an exhaustive list would be out of the scope of this work, we refer to~\cite{ConitzerW2016} for a comprehensive survey on manipulations and~\cite{FaliszewskiR2016,Lin2012} for an overview of known results in control and bribery.

Although computing a successful election attack is often computationally intractable, ongoing research still aims to develop algorithms to circumvent this obstacle. In complexity theory, the standard assumption for \textsf{NP}-hard problems is that we cannot expect an algorithm that simultaneously fulfills the following three criteria:
\begin{enumerate*}[label=\alph*)]
\item solves all instances,
\item gives optimal solutions, and
\item runs within polynomial time.
\end{enumerate*}
To explore the concrete boundaries of intractability, researchers often relax one of these requirements. 

The first criterion is the most natural candidate for exclusion, which is aligned with real-life observations that not all instances of the same problem are equally challenging. 
This principle also extends to the domain of elections.

In this line of research, we can investigate whether the shield is strong even in elections with a small number of candidates~\cite{ConitzerSL2007,HemaspaandraLM2016} or voters~\cite{BrandtHKS2013,ChenFNT2017}. However, most studies indicate that such restrictions provide substantial benefits only in very specific scenarios.

Additionally, we can explore the structured nature of voters' preferences. While attacks for single-peaked preferences can often be computed in polynomial time~\cite{Walsh07,FaliszewskiHHR2011,BrandtBHH2015}, other natural preference restrictions, such as nearly single-peaked preferences~\cite{Yang2018,Yang2020}, multi-peaked preferences~\cite{YangG2018}, single-crossing preferences~\cite{MagieraF2017}, group-separable preferences~\cite{FaliszewskiKO2022}, or spatial voting \cite{WuEKV2022}, there is no hope for any tractability for most voting rules. Furthermore, empirical evidence suggests that real-world elections rarely exhibit the structured characteristics necessary to maintain tractability~\cite{BredereckCW2016}.
All the above-mentioned restrictions of the input instances heavily employ the framework of parameterized complexity~\cite{CyganFKLMPPS2015,DowneyF2013,Niedermeier2006}.

Next, we can relax the requirement for an optimal solution, often leading to the exploration of approximation algorithms~\cite{Vazirani2001,WilliamsonS2011}. This research direction was initiated by~\citet{BrelsfordFHSS2008}, who presented both approximation algorithms and inapproximability results, focusing primarily on scoring protocols in the context of manipulation, control, and bribery attacks. This approach has been further pursued by various authors~\cite{ZuckermanPR2009,ZuckermanLR2011,FaliszewskiMS2021,KellerHH2019a,KellerHH2019b}.

Finally, we can drop the requirement on polynomial running time. In this direction, we try to decrease the base of the exponential function as much as possible. For example, the naive running time for the well-known \textsc{3-SAT} problem runs in $\mathcal{O}(2^n)$ time, where $n$ is the number of variables. On the contrary, the current fastest algorithm runs in $\mathcal{O}(1.34^n)$ time~\cite{MoserS2011}. A similar approach was also applied in the area of voting theory and influence of elections~\cite{KnopKM2020}. We would like to highlight here that many of these algorithms exploit techniques known from (integer) linear and quadratic programming.

For more comprehensive works on threats to the complexity shield established for the election, we refer the reader to the excellent survey of \citet{RotheS2013} and monographs by \citet{BrandtCELP2016} and \citet{Rothe2015}.

%%%%%%%%%%%%%%%%%%%%%%%%%%%%%%%%%%%%%%%%%%%%%%%%%%%%%%%%%%%%%%%%%%%%%%%%

\section{Basics of Quantum Computing}\label{sec:quantum}

Quantum computing represents a paradigm shift in computation that leverages the principles of quantum mechanics to process information. At the heart of quantum computing are quantum bits, or \emph{qubits}, which fundamentally differ from classical bits. In this section, we will introduce key concepts and principles of quantum computing.

More formally, in classical computing, the basic unit of information is a single bit, which can take values of $0$ or $1$. In contrast, qubits, the building blocks of quantum computation, can exist in multiple states simultaneously due to a quantum mechanics phenomenon known as \emph{superposition}. This means that a qubit can represent both 0 and 1 at the same time, which is a fundamental departure from classical bits. Superposition also allows us to perform many calculations significantly faster than in classical computers.

The second major underlying principle that differentiates quantum computing from classical computing is \emph{entanglement}. When two (or more) qubits become entangled, their properties become correlated regardless of how far they are from each other. That is, a change in the state of one qubit immediately affects the set of the other qubits. This property is of great importance in quantum communication.

The quantum counterparts of classical logical gates are \emph{quantum gates}. They operate on qubits and transform their states. Quantum gates are combined to form quantum circuits, an equivalent of digital circuits. Quantum circuits then execute quantum algorithms.

This high-level description provides a foundation for understanding quantum computing. For a more comprehensive introduction to quantum computing, we refer the interested reader to many monographs in the field, specifically those targeted at computer scientists, such as~\cite{NielsenC2016,YanofskyM2008,McMahon2008}.

Without elaborating on technical details, we would like to introduce at least one quantum algorithm that is of particular importance when dealing with \NPh problems, which generally contains exhaustive search as a subroutine, such as \textsc{3-SAT} or \textsc{Constraint Satisfaction Problem}~\cite{Ambainis2004,CerfGW2000}. Formally, we are given a function $f\colon\{0,\ldots, n-1\}\to\{0,1\}$ and our goal is to find $x$ such that $f(x) = 1$. In classical computation, the solution of this problem requires $\Omega(n)$ evaluations of the function. On the other hand, Grover's algorithm~\cite{Grover1996} can find the solution using $\mathcal{O}(\sqrt{n})$ evaluations\footnote{Later, it was proved that any quantum algorithm for this problem requires $\Omega(\sqrt{n})$ evaluations of $f$~\cite{BennettBBV1997}. That is, Grover's algorithm is asymptotically optimal.}. Consequently, Groover's algorithm gives quadratic speedup over the classical algorithms.

%%%%%%%%%%%%%%%%%%%%%%%%%%%%%%%%%%%%%%%%%%%%%%%%%%%%%%%%%%%%%%%%%%%%%%%%

%%%%%%%%%%%%%%%%%%%%%%%%%%%%%%%%%%%%%%%%%%%%%%%%%%%%%%%%%%%%%%%%%%%%%%%%
\section{A Case Study}\label{sec:cryptography}

For decades, the prevailing security paradigm mirrored the core principle of the complexity shield discussed earlier. In various forms, many cryptographic systems adhere to a general schema where communicating parties have access to encryption and decryption keys. With these keys, both encryption and decryption are computationally easy, while without them, decryption becomes nearly impossible due to the complexity of the underlying computational problem.

Arguably, the most well-known example of such a cryptographic scheme is the RSA algorithm~\cite{RivestSA1978}. The fundamental security principle underlying this algorithm is the factorization of large numbers. The current fastest factorization algorithm operates in $\exp(\log(n)^{1/3} \cdot \log\log(n)^{1/3})$ time, where $n$ is the number we need to factorize~\cite{LenstraL1993}. In practical applications, in order to decrypt captured messages, one would need to factorize numbers with at least $4096$ bits. Given the immense size of these numbers, it becomes evident that the algorithm would require centuries to produce a result; that is, the RSA algorithm can be assumed effectively unbreakable within a reasonable time frame.

The pivotal development in public key cryptography, of which the RSA algorithm is a prominent example, came with the introduction of Shor's algorithm in 1994~\cite{Shor1994}. This quantum algorithm has the remarkable capability to solve prime factorization problems in polynomial time~\cite{HarveyH2021,BeckamCDP1996}. In practical terms, this means that any adversary armed with a quantum computer containing a sufficient number of stable qubits can decrypt RSA-based communications, regardless of how we increase the bit length of encryption keys.

While there is currently no evidence to suggest that humans will possess the capability to construct a practical quantum computer with enough stable qubits in the near future, the landscape of cryptography has undeniably shifted. This transformative shift, described as a ``cryptopocalypse'' by some authors~\cite{Javed2014}, has paved the way for entirely new cryptographic approaches -- commonly referred to as (post-)quantum cryptography~\cite{Rogers2010,Grasselli2021,BernsteinBD2008}. Notably, these post-quantum cryptographic methods remain secure even in the presence of quantum computers.

%%%%%%%%%%%%%%%%%%%%%%%%%%%%%%%%%%%%%%%%%%%%%%%%%%%%%%%%%%%%%%%%%%%%%%%%

\section{Implications for Elections}\label{sec:threat}

In this section, we follow up on the previous sections and discuss possible threats and opportunities quantum computing represents for voting theory.

\subsection{Complexity Shield}

First of all, our central motivation in this paper is the complexity shield that may vanish due to the computational capabilities of quantum computers. On the other hand, this also opens avenues for further research. Namely, the area of quantum algorithms in the area of election manipulation is, to the best of our knowledge, still an uncharted territory. Also, quantum computation comes with its complexity classes, and it can be shown that some computational problems are computationally hard, even under the lens of quantum algorithmics. Such a result would be a very desirable step towards a \emph{quantum complexity shield}, which in our eyes is necessary for trustworthy elections.

\begin{opportunity}
    Develop new quantum algorithms for the realization of different attack vectors for different voting rules. Provide quantum guarantees for the complexity shield protecting elections.
\end{opportunity}

Furthermore, it is imperative to acknowledge that, as alluded to in \Cref{sec:shield}, numerous voting rules and, by extension, strategies to influence them can be naturally formulated as integer linear (or quadratic) programs~\cite{KnopKM2020}. Recent advances in quantum optimization~\cite{BrandaoS2017,Moll2018,YarkoniRBS2022,LiptonR2014} demonstrate the increased potency of quantum techniques in solving optimization programs, which exceed the capabilities of contemporary methods. Should quantum optimization prove to be effective within our investigative domain, it could potentially introduce additional fissures in the existing complexity shield.

\begin{opportunity}
    Investigate the implications of advances in quantum optimization for the computation of results of voting rules and associated election attacks.
\end{opportunity}

\subsection{New Axioms and Impossibility Theorems}

Second, it may be the case that the new era may need its own axioms specifically tailored to the quantum computing paradigm. These quantum axioms should address the unique properties and behaviors of quantum systems, offering a more comprehensive framework for understanding voting mechanisms in quantum environments.

\begin{opportunity}
    Extend the plethora of existing voting axioms with new ones that are grounded in the quantum world.
\end{opportunity}

In addition to rethinking voting axioms, the time has come to re-evaluate traditional voting rules in light of quantum computing. We anticipate that the quantum landscape may not only offer novel ways to express voters' preferences but also introduce entirely new definitions of voting rules that transcend our current understanding.

\begin{opportunity}\label{op:ballots}
    Define voting rules and ballot forms that take advantage of quantum computing.
\end{opportunity}

As we redefine axioms and voting rules for the quantum era, the final step in this transformative process is to revisit the foundational question of whether there exists a quantum analog of the Gibbard-Satterthwaite theorem. This inquiry is of paramount importance, as it has the potential to illuminate the constraints and possibilities of quantum voting systems.

By examining the combinations of axioms that are feasible within a quantum framework, we may unravel the intricate interplay between quantum principles and voting theory. The quest to determine whether a quantum counterpart to the Gibbard-Satterthwaite theorem exists(with respect to ballots and rules mentioned in Opportunity~\ref{op:ballots}) represents a crucial frontier in ensuring the integrity and resilience of elections in the quantum age.

\begin{opportunity}
    Investigate how the existing impossibility theorems behave under the quantum paradigm. Prove impossibility theorems tailored for quantum extension of standard voting theory.
\end{opportunity}

%%%%%%%%%%%%%%%%%%%%%%%%%%%%%%%%%%%%%%%%%%%%%%%%%%%%%%%%%%%%%%%%%%%%%%%%

\section{First Steps}\label{sec:first}

It should be pointed out that elections, in the widest meaning of the word, have already received some attention from the viewpoint of quantum computing. We identified the following two main lines of research.

The first line of research focuses mainly on the intriguing challenge of leader election within the area of distributed computing theory~\cite{TaniKM2012,KobayashiMT2014,Ganz2017}. In this context, the objective is to devise a mechanism by which a collective of autonomous entities, be they processors or agents endowed with computational capabilities, can designate a single entity as their leader. The designated leader then plays the critical role of coordinator and supervisor in ongoing tasks within the group.

The second line of research focuses on the security of electronic voting protocols~\cite{GuoFZ2016,GaoZGJH2019,LiJZL2021,WangLLYP2021,SunGDXY2022,WuSWCHDX2021}. 
These efforts undeniably contribute to the trustworthiness of the election, but they address the process of collecting and transmitting votes electronically, which is generally not the main focus of studies in social choice (with the notable exception of the manipulation of voting districts~\cite{LewenbergLR2017,LevL2019,EibenFPS2020} described earlier in this paper).

Although related research has made strides in various aspects of elections, the unique questions and opportunities we introduce here remain untouched.

%%%%%%%%%%%%%%%%%%%%%%%%%%%%%%%%%%%%%%%%%%%%%%%%%%%%%%%%%%%%%%%%%%%%%%%%
%\vspace{-0.16cm}

\section{Conclusions}\label{sec:conclusion}

In this work, we have explored the emerging threats and opportunities posed by the increasing prominence of quantum computing in the areas of voting, election manipulation, and computational social choice. Our central argument revolves around the fundamental premise that the security of voting rules is primarily based on the practical intractability of associated computational problems. However, the advent of quantum computers introduces hitherto unseen computational capabilities, potentially challenging the current complexity shield protecting different voting rules.

We firmly believe that, as a scientific community, we bear a small part of the responsibility for trustworthy and secure elections and voting rules. It is crucial to investigate how and whether current security guarantees can withstand the transformative influence of quantum computing. While ultimate confidence in election outcomes lies on the shoulders of politicians and representatives, robust scientific evidence of rules' security is indispensable for convincing arguments.

Although the threat to the elections' complexity shield may not be as immediate or severe as seen in cryptography with Shor's algorithm, we advocate initiating research in this direction sooner rather than later. It should be noted that neighboring fields such as artificial intelligence, machine learning, and multi-agent systems have already embarked on their quantum journey~\cite{Klusch2004,Klusch2005,YuYLW2022,Bausch2020,RicksV2003,WangYLC2021,LiWCW2021,HeidariGS2022,YunPK2023,BadescuO2021,Pastorello2023,YamasakiSSK2020,Castelvecchi2024}.

Lastly, we highlight that the Nobel Prize in Physics for 2022 was awarded to Aspect, Cluster, and Zeilinger ``for experiments with entangled photons, establishing the violation of Bell's inequalities and pioneering quantum information science''~\cite{NobelPrize2022}. From all perspectives, quantum computing stands as a formidable force, and we should not underestimate its relevance in the area of voting theory.

%%%%%%%%%%%%%%%%%%%%%%%%%%%%%%%%%%%%%%%%%%%%%%%%%%%%%%%%%%%%%%%%%%%%%%%%

%%% The acknowledgments section is defined using the "acks" environment
%%% (rather than an unnumbered section). The use of this environment 
%%% ensures the proper identification of the section in the article 
%%% metadata as well as the consistent spelling of the heading.

\begin{acks}
This work was co-funded by the European Union under the project ROBOPROX (reg. no. CZ.02.01.01/00/22\_008/0004590) and by the Grant Agency of the Czech Technical University in Prague, grant No. SGS23/205/OHK3/3T/18.
\end{acks}

%%%%%%%%%%%%%%%%%%%%%%%%%%%%%%%%%%%%%%%%%%%%%%%%%%%%%%%%%%%%%%%%%%%%%%%%

%%% The next two lines define, first, the bibliography style to be 
%%% applied, and, second, the bibliography file to be used.

\bibliographystyle{ACM-Reference-Format} 
\bibliography{references}

%%%%%%%%%%%%%%%%%%%%%%%%%%%%%%%%%%%%%%%%%%%%%%%%%%%%%%%%%%%%%%%%%%%%%%%%
	
\end{document}